\documentclass[aps,prl,twocolumn,showpacs,preprintnumbers,amsmath,amssymb,floatfix]{revtex4-2}
\usepackage{graphicx}
\usepackage{subfigure}
\usepackage{dcolumn}
\usepackage{bm}
\usepackage{color}
\usepackage{epstopdf}%
\usepackage{amsmath}%
\usepackage{amsfonts}%
\usepackage{amssymb}
\usepackage{bm}
\newcommand{\mb}{\mathbf}

\begin{document}
\title{Magnetochiral anisotropy-induced nonlinear Hall effect in spin-orbit coupled Rashba conductors}
\author{D. C. Marinescu}
\author{S. Tewari}
\affiliation{Department of Physics and Astronomy, Clemson University, Clemson, South Carolina 29634, USA}

\begin{abstract}
We theoretically predict the existence of a non-zero magnetochiral anisotropy-induced nonlinear Hall effect or a second harmonic Hall voltage transverse to an applied current in spin-orbit coupled Rashba conductors in the presence of an in-plane magnetic field $\mathbf{B}$. This is distinct from the Berry curvature dipole (BCD)-induced nonlinear Hall effect in systems with non-trivial bandstructure because the former requires broken time-reversal symmetry while the BCD-induced effect is non-zero even in time-reversal symmetric systems. This result is independent of the existence of other types of interactions, such as hexagonal warping or cubic Dresselhaus interactions.
%We calculate the
%effect by considering
%the second order correction to the non-equilibrium electron distribution function. We evaluate the latter by considering the local change
%effect by considering the local change in the single-particle energy due to the applied electric field $\mathbf{E}$ and expanding the electron distribution function perturbatively up to the quadratic order in the electric field.
We find that for $\mathbf{E}||\mathbf{B}$, the magnitude of the nonlinear Hall current flowing in a direction \textit{perpendicular} to the applied electric field is exactly $1/3$ of the magnitude of the non-linear rectification current responsible for magnetoresistance \textit{parallel} to the applied electric field obtained in the $\mathbf{E}\perp \mathbf{B}$ configuration. 
%Since the non-linear magnetoresistance parallel to the electric field has already been successfully observed in several Rashba systems, the magnetochiral anisotropy-induced nonlinear Hall effect proposed in this paper is guaranteed to be observable in the same systems.
%of coplanar electric and magnetic fields with
%$\mathbf{E}\perp \mathbf{B}$.
%For arbitrary angle $\theta$ between $\mathbf{E}$ and $\mathbf{B}$, both nonlinear Hall effect and the rectification current are non-zero, and we calculate their dependence on the in-plane angle $\theta$.  The temperature-independent current-nonlinear Hall effect exists for all values of the chemical potential, above or below the band crossing point (BCP), the energy where the two chiral conduction bands intersect. We predict that it would be visible in any 2D system with strong Rashba spin-orbit interaction, including III-V semiconductor quantum wells, and 3D Rashba-type polar semiconductors such as BiTeBr.
\end{abstract}

\date{\today}
\maketitle

{\textit{\textbf{Introduction:}}}
In electric charge Hall effect a transverse voltage is generated in response to a longitudinal current in conductors subjected to an out-of-plane magnetic field, magnetization, or non-trivial geometric properties of the band structure such as Berry curvature \cite{Luttinger1, Luttinger2, Jungwirth, Xiao}. In first order charge current response to an applied electric field ($\mathbf{j}_{\alpha}=\sigma_{\alpha \beta}\mathbf{E}_{\beta}$, where $\alpha,\beta =x,y,z$), the presence of a non-zero, nondissipative, Hall conductivity ($\sigma^{H}_{\alpha \beta}=-\sigma^{H}_{\beta \alpha}$) requires broken time reversal symmetry (TRS) \cite{Philippe, Onsager1, Onsager2}. In the nonlinear current regime ($\mathbf{j}_{\alpha}=\chi_{\alpha\beta\gamma}\mathbf{E}_{\beta}\mathbf{E}_{\gamma}$), a non-zero Hall voltage quadratically dependent on the applied electric field can result with or without broken TRS \cite{Spivak, Sodemann}. In TRS symmetric systems with broken spatial inversion (SI) symmetry, the dissipationless nonlinear Hall current can result from a non-zero first order moment of the Berry curvature, the so-called Berry curvature dipole (BCD), over the occupied bands \cite{Sodemann}. But when TRS and SI are simultaneously broken, a second class of nonlinear Hall response is allowed by symmetry, arising from magnetochiral anisotropy effect also known as nonreciprocal magnetotransport \cite{Tokura2018, Yokouchi, Yasuda}. Magnetochiral anisotropy implies that in non-centrosymmetric systems subjected to a magnetic field, the resistivity parallel and perpendicular to the applied current is different for the current flowing to the right ($+I$) and to the left ($-I$). This effect was predicted to lead to a nonlinear magnetoresistance effect \textit{parallel} to the applied electric field in Rashba semiconductors and was experimentally observed \cite{Rikken2005, Ideue2017, Li2021, Nagaosa2018, Pan2022}. However, the magnetochiral anisotropy-induced nonlinear Hall effect \textit{transverse} to the applied electric field was shown to vanish in Rashba systems \cite{He_PRL_2019, Dantas_PRB_2023} and was found to be non-zero only in exotic systems such as Weyl semimetals \cite{Yokouchi}, topological insulators \cite{Yasuda, He_PRL_2019}, and systems with cubic Dresselhaus interactions \cite{Dantas_PRB_2023}. In this paper, we demonstrate that just as nonlinear magnetoresistance, the nonlinear Hall effect due to magnetochiral anisotropy is non-zero even in Rashba systems. In addition to being theoretically important in placing the nonlinear Hall effect side by side with nonlinear magnetoresistance by demonstrating that both are non-zero and comparable in magnitude in Rashba semiconductors, our results also greatly enhance the scope of candidate systems in which the magnetochiral anisotropy-induced nonlinear Hall effect is likely to be experimentally observable.

%In this paper we theoretically predict the existence of a non-zero magnetochiral anisotropy-induced nonlinear Hall effect
%in a direction perpendicular to an applied electric field $\mathbf{E}$
%in simple spin-orbit coupled Rashba systems subjected to an in-plane $\mathbf{E}$ and $\mathbf{B}$ fields in a non-orthogonal configuration $\mathbf{E}\cdot \mathbf{B}\neq 0$. Previously, this effect was discussed only in the presence of additional interactions, such as hexagonal warping or cubic Dresselhaus terms \cite{He_PRL_2019, Dantas_PRB_2023}. Since the complementary effect - magnetochiral anisotropy-induced nonlinear magnetoresistance - has already been successfully observed \cite{Ideue2017, Li2021}, the nonlinear Hall effect predicted in this paper (without a non-zero BCD) is guaranteed to be experimentally observable in these systems.

Nonlinear transport effects in systems with Rashba spin-orbit interaction with coupling constant $\alpha$ rely on a second-order approximation of the non-equilibrium electron distribution function that is quadratic in the applied electric field. The parity of this non-equilibrium electron distribution function in the momentum space is even. When currents are calculated by summing electron velocities multiplied by the non-equilibrium distribution function, only the same parity component of the electron velocity is selected, which in these systems is proportional to the coupling constant\cite{yu,nagaosa}. This paradigm also underlies the phenomenology that occurs in the presence of coplanar electric and magnetic fields.  Since the magnetic field couples to the parallel projection of the electron spin through the Zeeman interaction (e.g., for magnetic field along $\hat{y}$, $\sim B_y \sigma_y$), while the spin in the same direction is coupled to the momentum in the orthogonal direction through Rashba SOI ($\sim \alpha (p_x \sigma_y - p_y \sigma_x)$ with $\alpha$ the Rashba coupling constant), the Zeeman coupling along $\hat{y}$ amounts in effect to a shift of the electron momentum along $x$ direction by $B_y/\alpha$. This is a constant shift of the electron momentum in the Rashba system perpendicular to the magnetic field that becomes macroscopically observable when multiplied by a distribution function of the same parity, i.e.,  $\sim(\mathbf{v}\cdot\mathbf{E})^2$. The shift in the electron momentum leads to the rectification effect that appears in orthogonal electric and magnetic fields, say $E_x$ and $B_y$, when an electric current linear in the magnetic field and quadratic in the electric field is induced, $j_{xyx}^{(2)} \sim B_yE_x^2$, parallel to the applied electric field. This is the so-called magnetochiral anisotropy-induced non-reciprocal magnetoresistance. %Recognized almost twenty years ago \cite{Rikken2005}, the existence of this non-linear magnetoresistance and its geometric dependence on the magnetic field has been detected experimentally in several systems with chiral energy bands, such as bulk Rashba semiconductors \cite{Ideue2017}, surface states of $\alpha$-GeTe \cite{Li2021}, and have been predicted even in superconductors \cite{Nagaosa2018}.

Here we highlight the complementary effect that occurs when the electric and magnetic fields are aligned, $E_x$ and $B_x$. In a geometry of coplanar but non-orthogonal electric and magnetic field with a non-zero $\mathbf{E}\cdot\mathbf{B}$), the magnetochiral anisotropy induces a nonlinear Hall effect generating a \textit{transverse} electric field along the $y$ direction, $E_{Hyxx}^{(2)}\sim B_xE_x^2$. 
%This effect is distinct from the non-linear Hall effect induced by a non-zero Berry curvature dipole \cite{Sodemann} because the former requires TRS and inversion breaking while the BCD-induced effect is non-zero even in TRS symmetric systems. %In particular,
%in contrast to the WSM with diverging Berry curvature and topological insulator heterostructures where this effect has been previously discussed,
%BCD vanishes in the spin-orbit coupled Rashba conductors with in-plane $\mathbf{E}, \mathbf{B}$ fields.
In particular, despite the BCD vanishing in the present system, a non-linear Hall current quadratic in the applied electric field is guaranteed to be observable due to magnetochiral anisotropy, with magnitude (for $\mathbf{E}||\mathbf{B}$) exactly $1/3$ of the magnitude of the corresponding nonlinear magnetoresistance (for $\mathbf{E}\perp\mathbf{B}$) which has already been successfully measured in Rashba systems \cite{Ideue2017, Li2021}.

The two complementary effects, nonlinear Hall and nonlinear magnetoresistance due to magnetochiral anisotropy, can be described quantitatively by merit coefficients $\gamma$ \cite{Yokouchi},
\begin{eqnarray}
  \gamma_x &=& -\frac{1}{B_y}\frac{\sigma_{2xx}}{(\sigma_{1})^2} \;,\label{eq:gamma-x} \\
  \gamma_y &=& -\frac{1}{B_x}\frac{\sigma_{2xy}}{\sigma_{1}^2}\;,\label{eq:gamma-y}
\end{eqnarray}
where $\sigma_1$ is the linear conductivity in the system, considered isotropic, while $\sigma_{2xx}$ and $\sigma_{2xy}$ are the second order conductivities. With $\gamma_{x,y}$ thus defined, the second order effects acquire a uniform description in terms of the equivalent electric fields, $E_{xxy} = \gamma_{x}j_x^2B_y/\sigma_1$ for nonlinear magnetoresistance and $E_{yxx} = \gamma_yj_x^2B_x/\sigma_1$ for the nonlinear Hall effect.

\begin{figure}\label{fig1}
\centering
\subfigure[]{\includegraphics[width=8 cm, height = 4.9cm,keepaspectratio]{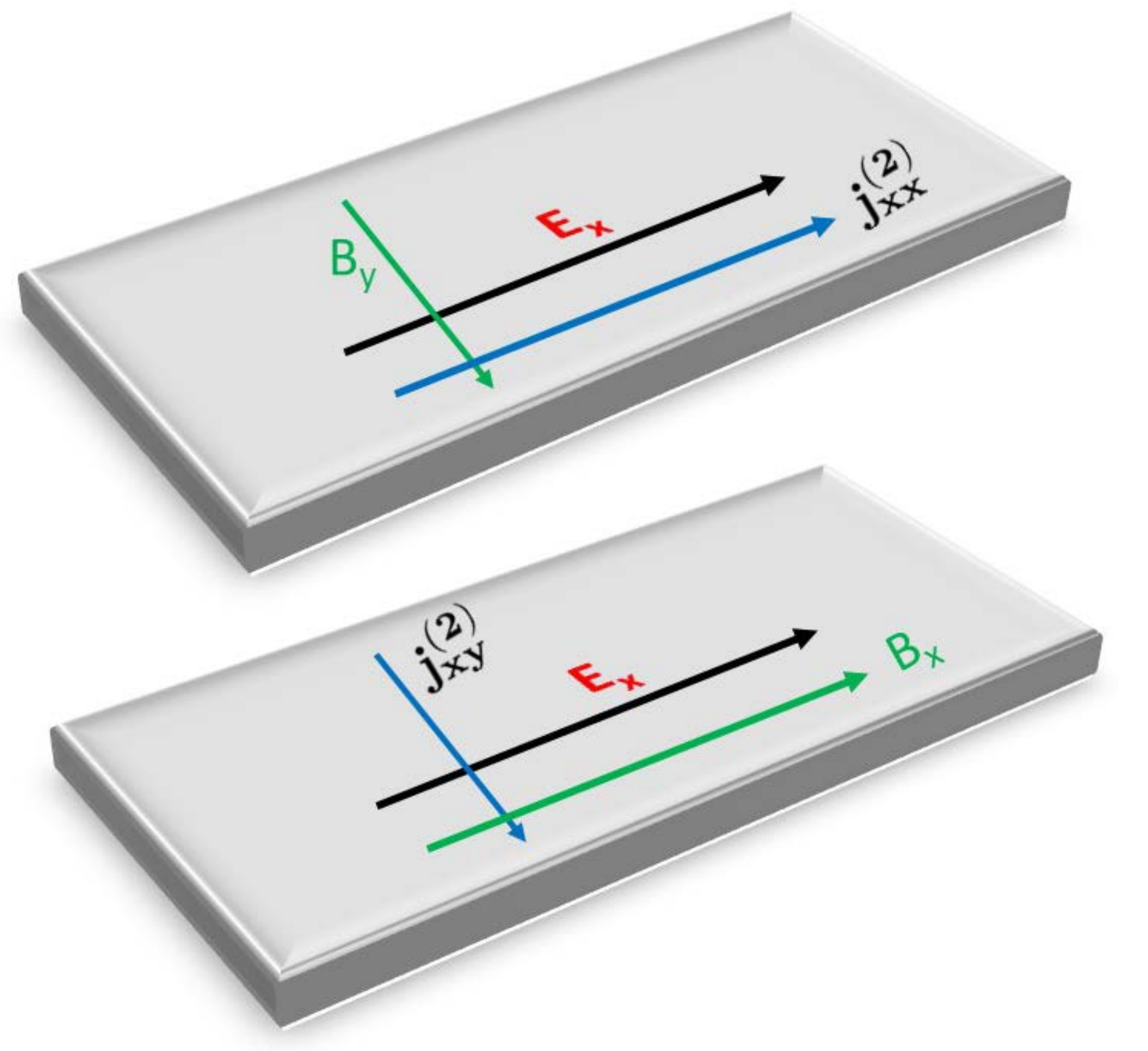}}\label{pic-a}
\subfigure[]{\includegraphics[width=4.2cm, height = 4.2cm, keepaspectratio]{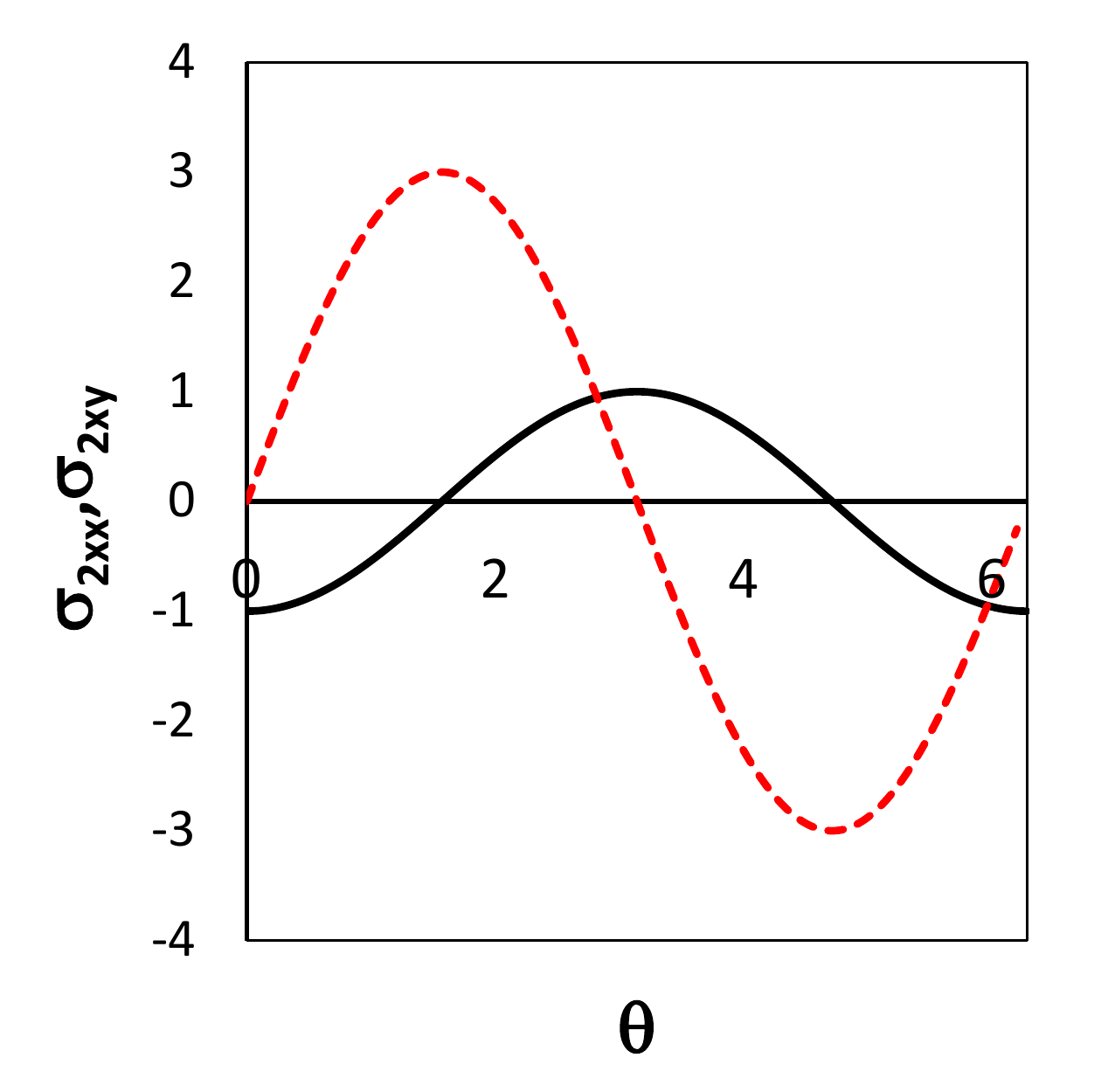}}\label{pic-c}
\subfigure[]{\includegraphics[width=4.2cm,height = 4.2cm,keepaspectratio]{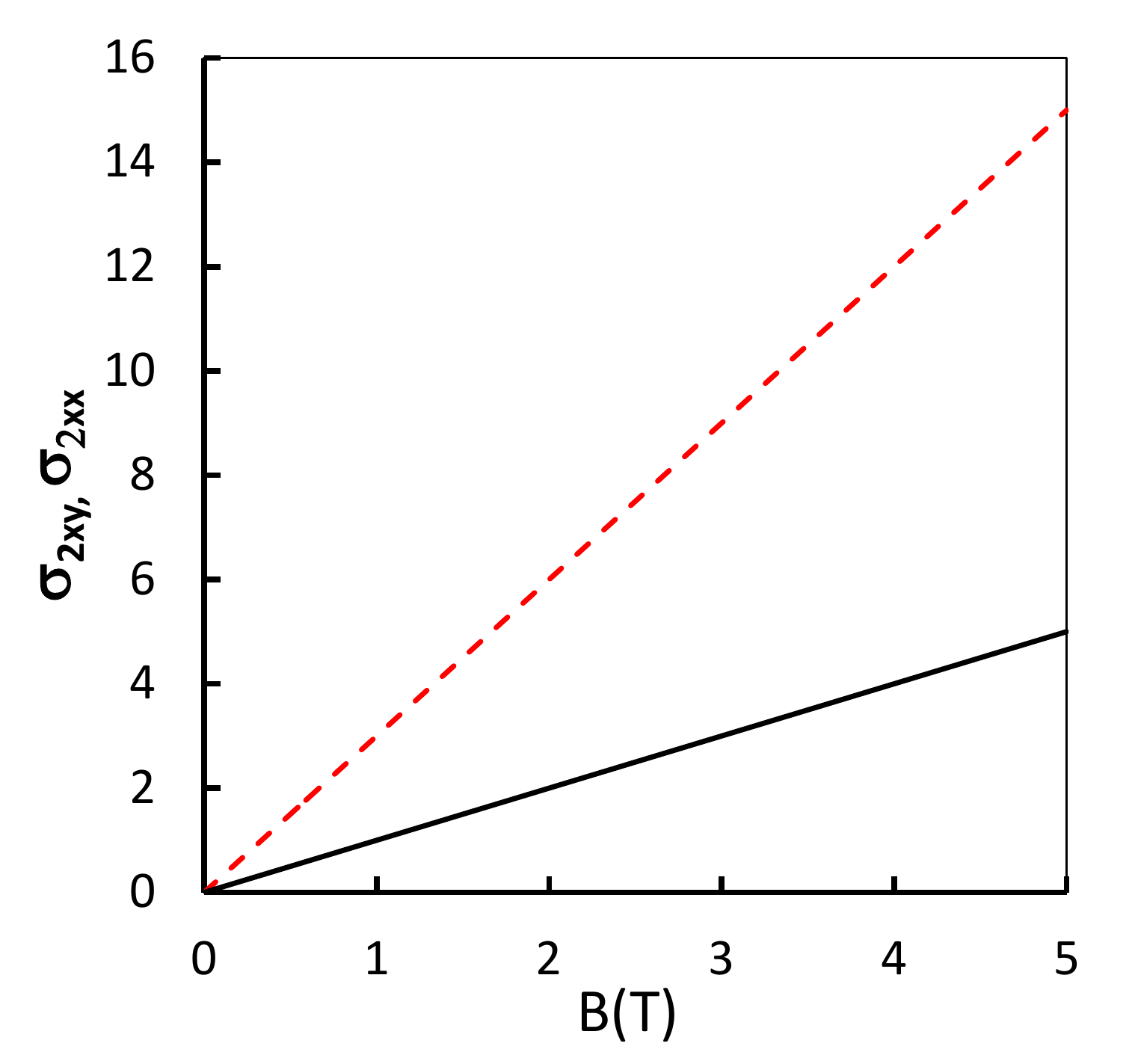}}\label{pic-d}
\caption{(a) Experimental configuration ($\mathbf{B}\perp \mathbf{E}$) for the detection of the maximum rectification current responsible for magnetoresistance parallel to the applied electric field, and the experimental geometry ($\mathbf{B}\parallel\mathbf{E}$) for measuring the  maximum second Harmonic Hall effect in a direction transverse to the applied electric field.
%(b) The band structure of a chiral Rashba system in the presence of a magnetic field.
(b)-(c)The angular and magnitude dependence, respectively, of the normalized $\sigma_{2xx}\sim B\sin\theta$, and $\sigma_{2xy}\sim B\cos\theta$ on the magnetic field.
The relative magnitudes satisfy, $\sigma_{2xy} = 1/3\sigma_{2xx}$ in (b) and (c).}
\label{fig1}
\end{figure}

In what follows, we first calculate the non-linear Hall current due to magnetochiral anisotropy in a 2D system with Rashba coupling. We demonstrate that this effect is non-zero for all values of the chemical potential, above or below the band crossing point (BCP) defined as the energy where the chiral bands intersect in the Rashba band structure.
We find that the magnitude of the Hall current transverse to the applied electric field is exactly $1/3$ of that of the rectification current responsible for the magnetoresistance parallel to the applied electric field for the same values of the magnetic field. The direct proportionality of the two currents is a consequence of the angular anisotropy of the Fermi surface in the presence of a magnetic field. Furthermore, we then extend our calculations to three dimensions and show that this relationship is maintained in bulk Rashba systems like BiTeBr, where the rectification current parallel to the applied electric field has recently been experimentally observed \cite{Ideue2017}. 
%BiTeBr is thus a three-dimensional polar semiconductor with Rashba spin-orbit coupling that provides an excellent testbed for the observation of the magnetochiral anisotropy-induced non-linear Hall effect proposed in this paper.

In Fig.~\ref{fig1}(a) we describe the experimental set-up for the measurement of the two nonlinear effects separately. On top, the applied electric and magnetic fields are orthogonal to each other, a geometry that can be used to measure the maximum nonlinear magnetoresistance parallel to the applied electric field. In the bottom panel of Fig.~\ref{fig1}(a) we show the geometry for measuring the maximum nonlinear Hall voltage proposed in this paper. For parallel configuration of the $\vec{E}$ and $\vec{B}$ fields, the Zeeman coupling due to the magnetic field leads to an effective shift of the momentum of the Rashba system \textit{transverse} to the applied electric field, which produces the nonlinear Hall effect. When the $\vec{E}$ and $\vec{B}$ fields are both in the $x-y$ plane but make an angle $\theta$ between them, the nonlinear Hall conductivity goes as $B\cos\theta$ and the nonlinear magnetoconductance depends of the magnetic field as $B\sin\theta$. In  Fig.~\ref{fig1} panels (b) and (c) we present the dependence of the normalized conductivities on the direction and magnitude of the applied magnetic field.

%This Hall effect is different from the non-linear Hall effect due to Berry curvature dipole \cite{nandy2017}, since the former requires time reversal symmetry breaking while the latter is non-zero even in the presence of time-reversal symmetry. Experimentally, they are distinguished by their dependence on the applied magnetic field, linear vs. quadratic in the case of the planar-Hall effect. \red{comments?}

{\textit{\textbf{Two-Dimensional electron gas with Rashba spin-orbit coupling:}}}
We begin by considering a 2D system with Rashba spin-orbit coupling, subjected to an in-plane electric field, $E_x$ and an in-plane magnetic field $\{B_x, B_y\}$. Later on we will extend the calculations to $3D$. In contrast to previous works, e.g., Ref.~\onlinecite{Ideue2017} and Ref.~\onlinecite{Pan2022} that included only the component of the magnetic field $B_y$ perpendicular to the applied electric field $E_x$ to investigate the nonlinear magnetoresistance parallel to $E_x$, in this work we consider an in-plane magnetic field forming an arbitrary angle $\theta$ with the electric field. It then follows that, in contrast to the previous papers, both components of the magnetic field $B_x$ and $B_y$ are part of the Rashba Hamiltonian in Eq.~(\ref{rashba-2D}) to describe the nonlinear Hall effect simultaneously with the magnetoresistance. Therefore, despite some superficial similarities of the calculations below with the equations contained in Ref.~\onlinecite{Pan2022}, the equations in the present work are significantly more involved and different from the previous works, so the final result in this paper is the transverse conductivity (Hall effect) perpendicular to the applied electric field, Eq.~(\ref{eq:sigma2}), while the result of the previous works \cite{Ideue2017, Pan2022} was nonlinear magnetoresistance parallel to the applied electric field.

The single-particle Hamiltonian of an electron of momentum $\mathbf{p} = \{p_x,p_y\}$, spin $\boldsymbol{\sigma} = \{\sigma_x,\sigma_y\}$, and effective mass $m$ is given by,
\begin{equation}
H_{2D} = \frac{p_x^2}{2m}+ \frac{p_y^2}{2m} + \alpha(p_x\sigma_y-p_y\sigma_x)-B_y\sigma_y-B_x\sigma_x\;,\label{rashba-2D}
\end{equation}
where $B_{x,y}$ designate the Zeeman splitting associated with the projections of the magnetic field along the two in-plane directions, $\hat{x}$ and $\hat{y}$, $g\mu_B B_{x,y}\sigma_y/2$ with $\mu_B$ the Bohr magneton and $g$ the effective gyromagnetic factor.
A canonical transformation, $p_x \rightarrow p_x + B_y/\alpha$, $p_y - B_x/\alpha$, leads to,
\begin{equation}\label{eq:rashba-2d}
H = \frac{1}{2m}\left(p_x + \frac{B_y}{\alpha}\right)^2 + \frac{1}{2m}\left(p_y -\frac{B_x}{\alpha}\right)^2 + \alpha (p_x\sigma_y-p_y\sigma_x)\;.
\end{equation}
The eigenstates of the Hamiltonian are,
\begin{equation}
|u_+\rangle = \frac{e^{i\mathbf{p}\cdot\mathbf{r}}} {\sqrt{2V}}\left(\begin{array}{c}1\\-ie^{i\phi}\end{array}\right)\;,\quad
|u_-\rangle  =  \frac{e^{i\mathbf{p}\cdot\mathbf{r}}}{\sqrt{2V}}\left(\begin{array}{c}-ie^{-i\phi}\\1\end{array}\right)\;, \label{eq:eigenstates}
\end{equation}
with $\tan \phi = -p_x/p_y$. The corresponding eigenvalues are,
\begin{eqnarray}
E_\xi & = & \frac{1}{2m} \left(p + \xi m\alpha + \frac{\Delta B}{\alpha}\right)^2 - \frac{1}{2m}\left(\xi m\alpha + \frac{\Delta B}{\alpha}\right)^2\nonumber\\
& + & \frac{B^2}{2m\alpha^2}\;,\label{eq:eigenvalues}
\end{eqnarray}
where $\Delta B = B_y\cos\varphi - B_x\sin\varphi$ and $p = \sqrt{p_x^2 + p_y^2}$ and $\xi = \pm 1$.

The two rotational paraboloid Fermi surfaces intersect at $\mathbf{p} = 0$, when $E_{BCP} = \frac{B^2}{2m\alpha^2} $. This energy value denotes the band crossing point (BCP), henceforth considered the origin of the 2D energy. Henceforth, this value will be set to $0$, since the Zeeman splitting $B$ is much lower than the Rashba energy, $m\alpha^2/2$ and quadratic terms in $B$ are neglected.

Equation $E_\xi = \epsilon$ generates the expressions of the momenta as a function of energy.  When $ -\frac{m\alpha^2}{2} + B\le \epsilon \le 0$, a condition that allows a complete encircling of the Fermi surface, both solutions, indexed by $\eta = \pm 1$ correspond to $\xi = -1$,
\begin{equation}\label{eq:fmom-below}
p_{-}^{\eta} = m\alpha -\frac{\Delta B}{\alpha} +\eta \sqrt{\left(\xi m\alpha + \frac{\Delta B}{\alpha}\right)^2 + 2m\epsilon} \;.
\end{equation}

For $\epsilon\ge 0$, momenta are associated with both minibands, $\xi = \pm 1$,
\begin{equation}\label{eq:fmom-above}
p_\xi = \sqrt{\left(\xi m\alpha + \frac{\Delta B}{\alpha}\right)^2 + 2m\epsilon} - \xi m\alpha - \frac{\Delta B}{\alpha}\;.
\end{equation}
When $\epsilon = \mu$, the effective 2D chemical potential, Eqs.~(\ref{eq:fmom-below}) and (\ref{eq:fmom-above}) define the Fermi momenta.

{\textit{\textbf{The nonlinear Boltzmann transport formalism and expansion in powers of E:}}} In the presence of an electric field, the electron distribution function $f_\mathbf{p}$
can be written as a power series expansion in $\mathbf{E}$:
\begin{equation}f_\mathbf{p} = f_\mathbf{p}^0 + \delta f_\mathbf{p}^{(1)} + \delta f_\mathbf{p}^{(2)}\end{equation}
Here $f^0(\epsilon_\mathbf{p}) = \left[e^{\beta(\epsilon_{\mathbf{p}}-\mu)} + 1\right]^{-1}$ is the equalibrium Fermi distribution function, while $\delta f_\mathbf{p}^{(1)}$ and $\delta f_\mathbf{p}^{(2)}$ are the non-equilibrium corrections proportional to $E_x$ and $E_x^2$ respectively.
$\delta f_\mathbf{p}^{(1)}$ is the solution to the Boltzmann transport equation \cite{ashcroft},
\begin{equation}
\delta f_{\mathbf p}^{(1)} = e\tau\mathbf{v}_{\mathbf{p}}\cdot \mathbf E \frac{df^0}{d\epsilon} \;,\label{eq:ef1}
\end{equation}
where $v_{\mathbf{p}} = \nabla_{\mathbf p}\epsilon_\mathbf{p}$ is the electron velocity.
To obtain the sesecond-orderistribution function, we use a semi-classical approximation of the local perturbation induced by the external fields on the distribution function.
 Thus, the addition of an electrostatic potential $V(\mb r) = -\mb E\cdot \mb r$, modifies locally the electron energy to $\tilde{\epsilon}_{\mb p} = \epsilon_{\mb p} + e\mb E\cdot\mb r$, a change considered weak with respect to the Fermi energy.
 In this approximation, the second order distribution function becomes,
\begin{equation}
\delta f_{\mb p}^{(2)} = \frac{1}{2k_BT}\left(e\tau\mb E\cdot\mb v_{\mb p}\right)^2\tanh \frac{\epsilon_{\mb p} - \epsilon_F}{2k_BT}\left(-\frac{\partial f_{\mb p}^0}{\partial \epsilon_{\mb p}}\right)\;. \label{eq:ef2}
\end{equation}

The single particle velocities are, from Eq.~(\ref{eq:eigenvalues}),
\begin{eqnarray}\label{eq:velocities}
  v_x = \frac{\partial E_\xi}{\partial p_x} &=& \frac{1}{m}\left(p\cos\varphi + \frac{B_y}{\alpha}\right) + \xi \alpha\cos\varphi\;, \\
  v_y =\frac{\partial E_\xi}{\partial p_y}  & = & \frac{1}{m}\left(p\sin\varphi - \frac{B_x}{\alpha}\right) + \xi \alpha\sin\varphi\;.
\end{eqnarray}

As a function of energies, velocities are obtained by inserting Eqs.~(\ref{eq:fmom-below}) and (\ref{eq:fmom-above}) in the above equations:
\begin{itemize}
  \item $-\frac{m\alpha^2}{2}\le \epsilon\le 0$
  \begin{eqnarray}
  v_x &=& \eta\frac{p_1}{m}\cos\varphi - \frac{\Delta B\cos\varphi}{m\alpha} +\frac{B_y}{m\alpha}\;, \\
  v_y &=& \eta\frac{p_1}{m}\sin\varphi - \frac{\Delta B\sin\varphi}{m\alpha} -\frac{B_x}{m\alpha}\;,
\end{eqnarray}
where $p_1 = \sqrt{\left(m\alpha - \frac{\Delta B}{\alpha}\right)^2 + 2m\epsilon}$.
  \item $\epsilon\ge 0$
\begin{eqnarray}
  v_x &=& \frac{p_1}{m}\cos\varphi - \frac{\Delta B\cos\varphi}{m\alpha} +\frac{B_y}{m\alpha}\;, \\
  v_y &=& \frac{p_1}{m}\sin\varphi - \frac{\Delta B\sin\varphi}{m\alpha} -\frac{B_x}{m\alpha}\;,
\end{eqnarray}
where $p_1 = \sqrt{\left(\xi m\alpha + \frac{\Delta B}{\alpha}\right)^2 + 2m\epsilon}$.
\end{itemize}
In the following considerations it is also important to introduce the SOI-induced corrections on the relaxation times, evaluated at the Fermi energy $\mu$ \cite{Verma2019, Kapri2021, Pan2022}. This is a result of the SOI effect on the state functions Eqs.~ (\ref{eq:eigenstates}) that determine the scattering matrix element, as well as on the density of states. Thus,
for $\mu<0$,
\begin{equation}
\frac{\hbar}{\tau_\eta} = \frac{p_0}{m\alpha}\left(1-\eta \frac{p_0}{2m\alpha}\right)\;, \label{eq:tau-eta}
\end{equation}
while for $\mu>0$,
\begin{equation}
\frac{\hbar}{\tau_\xi} = \frac{\hbar}{\tau}\left(1+\xi\frac{m\alpha}{2p_0}\right)\;,\label{eq:tau-xi}
\end{equation}
with
\begin{equation}
p_0 = \sqrt{(m \alpha)^2 + 2m \mu} \;, \label{eq:p0}
\end{equation}
%\red{Eq.~(\ref{eq:tau-eta}) corrects an inadvertent error in \cite{pan2022} and thus modifies the $\mu$ dependence of all calculated currents, including the rectification one \cite{inoue2017, pan2022} for $\mu \le 0$. Should be kept like this or buried as a comment in the reference list?\cite{*[{prepended text}][{appended text}]pan2022}.}

 {\textit{\textbf{Second order transverse currents and nonlinear planar Hall conductivity:}}}
Because $\delta f^{(2)}_\mathbf{p}$ in Eq.~(\ref{eq:ef2}) is proportional to $\tanh(\epsilon_\mathbf{p} - \mu)$ which cancels at $\mu$, the summation algorithm for the currents uses the Sommerfeld expansion.
\begin{itemize}
\item{{$-\frac{m\alpha^2}{2}\le \mu\le 0$}}\\
The current is given by
\begin{eqnarray}
& & j^{(2)}_{y,<} = -e\sum_{p\in[p_{-}^{-},p_-^+]}v_{x\eta}\delta f^{(2)}_\mathbf{p} = -\frac{e^3\tau^2 E_x^2}{48\hbar^2} \nonumber\\
& \times & \sum_\eta\eta \tau_\eta^2 \int_0^{2\pi}\frac{d}{d\epsilon}\left[\left(p_-^{\eta} \frac{dp_-^\eta}{d\epsilon}v_yv_x^2\right)\right]_{\mu}d\varphi\;.
\end{eqnarray}
When the kernel is linearized in $B$, we obtain,
\begin{eqnarray}\label{jbelow}
& & j_{y,<}^{(2)} =  - \frac{e^3E_x^2B_x}{48\hbar^2p_0}\sum_\eta \tau_\eta^2
 \int_0^{2\pi}\left(\frac{3}{2}\sin^22\varphi - \cos^2\varphi\right)d\varphi\nonumber\\
 & = & -\frac{e^3\pi\tau^2 E_x^2}{48\hbar^2}\frac{B_xp_0}{(m\alpha)^2}\;.
 \end{eqnarray}

\item{{$\mu\ge 0$}}\\
The current is,
\begin{eqnarray}
& & j_{y,>}^{(2)} = - e\sum_{\mathbf p, \xi}v_{x\xi}\delta f^{(2)}_\mathbf{p}\nonumber\\
 &&= -\frac{e^3E_x^2}{48\hbar^2}\sum_\xi\tau_\xi^2\int_0^{2\pi}\frac{d}{d\epsilon}\left[p_\xi\frac{dp_\xi}{d\epsilon} v_{y\xi} v_{x\xi}^2\right]_{\mu}d\varphi\;,
 \end{eqnarray}
 where we explicitly consider the chiral dependence of the relaxation time, Eq.~(\ref{eq:tau-xi}).

In first order in $B$, we obtain,
\begin{eqnarray}\label{jabove}
& & j_{y,>}^{(2)} = - e\sum_{\mathbf p, \xi}v_{x\xi}\delta f^{(2)}_\mathbf{p} \nonumber\\
& = & -\frac{e^3\pi E_x^2}{48\hbar^2}\sum_\xi\tau_\xi^2\left(-\xi\frac{B_x}{p_0}\right)\int_0^{2\pi}\left(\frac{3}{2}\sin^22\varphi - \cos^2\varphi\right)d\varphi\nonumber\\
 & = & -\frac{e^3\pi\tau^2 E_x^2}{96\hbar^2}\sum_\xi\left(1-\xi\frac{m\alpha}{p_0}\right)\left(-\xi\frac{B_x}{p_0}\right) \nonumber\\
 &= &-\frac{e^3\pi E_x^2}{48\hbar^2}\frac{m\alpha B_x}{p_0^2}\;.
 \end{eqnarray}
\end{itemize}
%\begin{figure}[ht]
%\centering
%\includegraphics[scale=.4]{fig2.pdf}
%\caption{The 2D second order transverse conductivity in Eq.~(\ref{eq:sigma2}).}
%\label{fg2}
%\end{figure}

%\begin{figure}
%\centering
%{\includegraphics[width=6.5cm]{fig2.pdf}}
%\caption{ The normalized $\sigma_{2xy}$ and $\sigma_{2xx}$ as functions of the chemical potential. $\sigma_{2xy} = 1/3\sigma_{2xx}$. The two quadratic %conductivities are continuous across the BCP. }
%\end{figure}
Using Eqs.~(\ref{jbelow}) and (\ref{jabove}) we write the expressions for the transverse conductivity perpendicular to the applied electric field (Hall conductivity),
 \begin{equation}\label{eq:sigma2}
 \sigma_{2xy} = \left\{\begin{array}{ll}
                         -\frac{e^3\pi\tau^2}{48\hbar^2}\frac{B_x}{\sqrt{(m\alpha)^2 + 2m\mu}} \left(1+\frac{2\mu}{m\alpha^2}\right) & \mu\le 0 \\
                         -\frac{e^3\pi\tau^2}{48\hbar^2}\frac{m\alpha B_x}{(m\alpha)^2 + 2m\mu}\, & \mu \ge 0
                       \end{array}\right.
                       \end{equation}
These values represent exactly $1/3$ of the rectification current leading to nonlinear magnetoresistance parallel to the electric field \cite{Pan2022}, $\sigma_{2xy} = 1/3\sigma_{2xx}$.
In an extremely simplified picture, the $1/3$ ratio between the longitudinal and transverse conductivities originates in the coefficients of the Taylor expansions of the electron velocities in the magnetic energy: the longitudinal current depends on $v_x^3\sim (p_0^2-mB_y)^{3/2}\sim p_0^3-\frac{3}{2} mB_y p_0,$ with $p_0$ the equilibrium momentum in Eq.~(\ref{eq:p0}), while the transverse current is given by $v_y v_x^2\sim (p_0^2+mB_x)^{1/2} p_0^2 \sim p_0^3+\frac{1}{2} B_x mp_0.$ Of course, these expressions ignore the angular dependence of the velocities which needs to couple into similar terms proportional with the Rashba interaction to survive the angular integration, along with the chiral symmetry requirement.

\textit{\textbf{Extension to 3D:}} The above results can be easily extended to three dimensional Rashba semiconductors, whose Hamiltonian is obtained from Eq.~(\ref{rashba-2D}),
\begin{equation}\label{rashba-3d}
  H_{3D} = \frac{p_z^2}{2m_\parallel} + H_{2D}\;,
\end{equation}
where $m_\parallel$ is the effective mass along the $\hat{z}$ direction.
In that case, the 2D chemical potential $\mu$ is rewritten as $\mu = \epsilon_F - \frac{p_z^2}{2m_\parallel}$, where $\epsilon_F$ is the 3D chemical potential. With this substitution, the 3D currents are obtained through integration over $p_z$,
\begin{equation}\label{3dcurrent}
j_{3D} = \frac{1}{2\pi\hbar}\int_{p_{z\min}}^{p_{z\max}} j_{2D}dp_z\;,
\end{equation}
where $p_{z\min}$ and $p_{z\max}$ are determined by the position of $\epsilon_F$ in respect with the band crossing point, considered zero as before.
For
 $-m\alpha^2/2 <\epsilon_F<0$, contributions to the current come only from the states within Region I ($-m\alpha^2/2 <\mu<0$) in Fig.~1 (b). With $p_{z\min} = 0$ and $p_{z\max} = \sqrt{2m_\parallel\epsilon_F + m_\parallel{m\alpha^2}}$, Eq.~(\ref{3dcurrent}) generates,
\begin{equation}\label{eq:sigma2-3d}
  (\sigma_{2xy})_{3D} = -\frac{e^3\pi \tau^2 E_x^2}{384\hbar^3}\sqrt{\frac{m_\parallel}{m}}\left(1+\frac{2\epsilon_F}{m\alpha^2}\right)\;.
  \end{equation}
For $\epsilon_F >0$, the current is produced by states in Region I, when $p_{z\min} = \sqrt{2m_\parallel \epsilon_F}$ and $p_{z\max} = \sqrt{2m_\parallel\epsilon_F + m_\parallel{m\alpha^2}}$, and states from Region II, when $p_{z\min} = 0$ and $p_{z\max} = \sqrt{2m_\parallel\epsilon_F}$. With these,
\begin{eqnarray}\label{sigma2-3d-positive}
  & &(\sigma_{2xy})_{3D} = \frac{e^3\tau^2B_y}{192\pi\hbar^3}\sqrt{\frac{m_\parallel}{m}}\left[\left(\frac{2\epsilon_F}{m\alpha^2} +1\right)\tan^{-1}\sqrt{\frac{m\alpha^2}{2\epsilon_F}}\right. \nonumber\\
  &-& \left. \sqrt{\frac{1}{\frac{2\epsilon_F}{m\alpha^2}+1}}\right]\nonumber\\
  & &+ \frac{\pi e^3\tau^2}{96\hbar^3}\sqrt{\frac{m_\parallel}{m}}\frac{m\alpha B_y}{\sqrt{2m\epsilon_F + (m \alpha)^2}}\ln\left(\frac{\sqrt{\frac{m\alpha^2}{2\epsilon_F}+1}+1}{\sqrt{\frac{m\alpha^2}{2\epsilon_F}+1}-1}\right)\nonumber\\\;.
\end{eqnarray}
$\sigma_{2xy}$ is continuous across the BCP point, $\epsilon_F = 0$, and cancels when $\alpha \rightarrow 0$. Since the kernel of the integral (\ref{3dcurrent}) scales with $j_{2D}$, the rectification conductivity satisfies $(\sigma_{2xx})_{3D} = 3(\sigma_{2xy})_{3D}$.

Numerical estimates indicate that the magnetochiral anisotropy-induced nonlinear Hall effect should be observable in III-V quantum wells. Using the typical numbers for GaAs quantum well with a particle density $n = 7\times 10^{15}$m$^{-2}$, effective mass $m=0.067m_e$, Rashba coupling $\alpha = 1.6\times 10^{4}$m/s, effective gyromagnetic factor $g = -0.44$, from Eq.~(\ref{eq:gamma-y}) we obtain a value for $\gamma_y = 1.3\times 10^{-8}$T$-1$A$^{-1}$m$^{2}$ (the conductivity $\sigma_1 = ne^2\tau/m$). In bulk Rashba systems, this effect should also be experimentally observable since the corresponding coefficient for the nonlinear magnetoresistance $\gamma_x$, which is of the same order of magnitude, has already been measured \cite{Ideue2017, Li2021}.

{\textit{Conclusions:}}
 We show that, in the presence of parallel electric and magnetic fields, a magnetochiral anisotropy-induced transverse nonlinear Hall current is generated in a 2D spin-orbit coupled Rashba system. 
 %This effect is distinct from the Berry curvature dipole-induced nonlinear Hall effect in systems with nontrivial bandstructure \cite{Sodemann} because the former requires TRS breaking while the latter is non-zero even in time-reversal symmetric systems. 
 This effect is temperature-independent and exists for all values of the chemical potential. Its magnitude (for $\mathbf{E}||\mathbf{B}$) is theoretically calculated to be $1/3$ of the longitudinal second order rectification magnetoconductivity for $\mathbf{E}\perp\mathbf{B}$ for the same value of the applied magnetic field. This is a universal result that persists in 3D, a reflection of the spatial anisotropy of the Fermi surface that determines the angular dependence of the energy integrals. Longitudinal rectification magnetoresistance has been successfully observed in several Rashba systems, e.g., polar semiconductor BiTeBr \cite{Ideue2017}, surface states of
$\alpha$-GeTe \cite{Li2021}, and is even proposed in superconductors \cite{Nagaosa2018}. These systems and 2D quantum wells with Rashba-type SOI which are widely available are therefore prime candidates for experimentally observing magnetochiral anisotropy-induced nonlinear Hall effect predicted in our work.

\textit{Acknowledgement:} S.T. acknowledges support from ARO Grant No: W911NF2210247 and ONR Grant No: N00014-23-1-2061.

\end{document}